\documentclass{article}

\usepackage{arxiv}

\usepackage[utf8]{inputenc} 
\usepackage[T1]{fontenc}    
\usepackage{hyperref}       
\usepackage{url}            
\usepackage{booktabs}       
\usepackage{amsfonts}       
\usepackage{nicefrac}       
\usepackage{microtype}      
\usepackage{lipsum}		
\usepackage{graphicx}
\usepackage{natbib}
\usepackage{doi}
\usepackage{graphicx}%
\usepackage{multirow}%
\usepackage{amsmath,amssymb,amsfonts}%
\usepackage{amsthm}%
\usepackage{mathrsfs}%
\usepackage[title]{appendix}%
\usepackage{xcolor}%
\usepackage{textcomp}%
\usepackage{manyfoot}%
\usepackage{booktabs}%
\usepackage{algorithm}%
\usepackage{algorithmicx}%
\usepackage{algpseudocode}%
\usepackage{listings}%
\usepackage{tabularx}
\usepackage{float}
\usepackage{array}
\usepackage{caption}
\usepackage{mdframed}
\usepackage{threeparttable}

\title{Tales of the 2025 Los Angeles Fire: Hotwash for Public Health Concerns in Reddit via LLM-Enhanced Topic Modeling}

\author{
\bf  Sulong Zhou$^{1}$, 
\bf  Qunying Huang$^2$, 
Shaoheng Zhou$^3$, 
Yun Hang$^4$\textsuperscript{,*}, 
Xinyue Ye$^2$\textsuperscript{,*},\\
\bf Aodong Mei$^4$, 
\bf Kathryn Phung$^4$, 
\bf Yuning Ye$^1$, 
\bf Uma Govindswamy$^4$, 
\bf Zehan Li$^5$ \\[2pt] 
  $^1$Department of Landscape Architecture and Urban Planning \\
     \& Urban Artificial Intelligence Lab, Texas A\&M University, College Station, TX, 77840\\ 
  $^2$Geography, University of Wisconsin-Madison, Madison, WI, 53715\\
  $^3$Google, Mountain View, CA, 94035\\
  $^4$Department of Environmental and Occupational Health Sciences,\\ 
      School of Public Health, University of Texas Health Science Center at Houston, Houston, TX, 77030\\
  $^5$ McWilliams School of Biomedical Informatics, \\ University of Texas Health Science Center at Houston, Houston, TX, 77030\\
}

\renewcommand{\shorttitle}{Tales of the 2025 Los Angeles Fire}

\hypersetup{
pdftitle={A template for the arxiv style},
pdfsubject={q-bio.NC, q-bio.QM},
pdfauthor={Sulong Zhou, \textit{et al}},
pdfkeywords={social media, topic modeling, public health},
}

\begin{document}
\maketitle

\renewcommand{\thefootnote}{\fnsymbol{footnote}}  
\footnotetext[1]{Corresponding authors: yun.hang@uth.tmc.edu and xinyue.ye@gmail.com.}
\footnotetext[2]{Contributing authors: sulong.zhou@tamu.edu, qhuang46@wisc.edu, shawnzhou@google.com, aodong.mei@uth.tmc.edu, kathryn.phung@uth.tmc.edu, yuning.ye@tamu.edu, uma.m.govindswamy@uth.tmc.edu, zehan.li@uth.tmc.edu}

\begin{abstract}
Wildfires have become increasingly frequent, irregular, and severe in recent years. Understanding how affected populations perceive and respond during wildfire crises is critical for timely and empathetic disaster response. Social media platforms offer a crowd-sourced channel to capture evolving public discourse, providing hyperlocal information and insight into public sentiment. This study analyzes Reddit discourse during the 2025 Los Angeles wildfires, spanning from the onset of the disaster to full containment. We collect 385 posts and 114,879 comments related to the Palisades and Eaton fires. We adopt topic modeling methods to identify the latent topics, enhanced by large language models (LLMs) and human-in-the-loop (HITL) refinement. Furthermore, we develop a hierarchical framework to categorize latent topics, consisting of two main categories, Situational Awareness (SA) and Crisis Narratives (CN). The volume of SA category closely aligns with real-world fire progressions, peaking within the first 2–5 days as the fires reach the maximum extent. The most frequent co-occurring category set of public health and safety, loss and damage, and emergency resources expands on a wide range of health-related latent topics, including environmental health, occupational health, and one health. Respectively, these topics discuss air quality and water contamination, health risks of firefighters and frontline journalists, long-term ecological health, and equitable access for vulnerable populations such as pregnant women and the elderly. Grief signals and mental health risks consistently accounted for 60\% and 40\% of CN instances, respectively, with the highest total volume occurring at night. This study contributes the first annotated social media dataset on the 2025 LA fires, and introduces a scalable multi-layer framework that leverages topic modeling for crisis discourse analysis. By identifying persistent public health concerns, our results can inform more empathetic and adaptive strategies for disaster response, public health communication, and future research in comparable climate-related disaster events.
\end{abstract}


\section{Introduction}
Wildfires have intensified in intensity, severity, and duration on a global scale, largely due to the effects of climate change and increased human activity \cite{swain2025hydroclimate, mansoor2022elevation,yue2021assessing}. These unplanned fire events result in widespread structural damage, economic losses, environmental degradation, and serious public health threats to affected communities \cite{bowman2017human, d2022wildfire, rappold2017community,wang2016spatial}. The health risks associated with wildfires extend far beyond the immediate fire zones, encompassing not only physical and mental health concerns but also disparities in access to healthcare resources. Epidemiological and toxicological evidence links exposure to wildfire-derived fine particulate matter (PM2.5) to a range of symptoms, including respiratory and ocular irritation, and to adverse health outcomes such as asthma, cardiovascular diseases, negative birth outcomes, and skin inflammation  \cite{zhang2025respiratory, lei2024wildfire}. In addition, psychological impacts—such as anxiety, depression, and trauma-related disorders can result from smoke exposure, fear of loss, and evacuation-related stress \cite{lei2024wildfire, wettstein2024psychotropic}. These mental health effects may persist long after fires are contained and can exacerbate physical health issues, particularly among vulnerable populations including children, older adults, pregnant individuals, those with preexisting cardiopulmonary conditions, and individuals from lower socioeconomic backgrounds \cite{lei2024wildfire, rappold2017community}. 

When wildfires occur during irregular seasons and communities are unprepared, the consequences can be even more severe. In California's Mediterranean climatic region, January has historically been a low-probability period for large-scale wildfires \cite{dong2022season}. However, a hydroclimate whiplash 
\cite{homann2022linked, swain2025hydroclimate}, driven by low humidity and hurricane-force Santa Ana winds, has induced abrupt changes from protracted wet periods to extreme dryness. This shift triggered the most devastating winter wildfires in the history. As of January 31, 2025, a total of 28 wildfires had burned 57,636 acres, resulted in 29 fatalities (pending coroner confirmation), destroyed 16,244 structures \cite{calfire_incidents}, and forced the evacuation of approximately 200,000 residents \cite{StellohEtAl2025}. Preliminary data indicate that the Eaton and Palisades fires in Los Angeles rank second and third in structural damage, and fifth and ninth in fatalities, respectively, though these figures remain provisional \cite{calfire_statistics}. Despite the Storm Prediction Center \cite{SPC_FW2025}, National Interagency Fire Center \cite{NIFC_outlook}, and the local National Weather Service \cite{NWS_RFWLOX2025} had accurately forecasted the windstorm and heightened fire risk, the actual spread and scale of wildfires far exceeded expectations. 

Amid rapidly evolving wildfire conditions, many individuals have turned to social media platforms—such as Twitter and Reddit—for hyperlocal information, health-related resources, and mutual aid \cite{slavkovikj2014review, lever2022social, ma2024investigating}. While Twitter has been extensively studied in the context of disaster response, Reddit’s role in collective information sharing and community support remains comparatively underexplored \cite{chen2019online}. Traditional data sources, including clinical records and post-disaster surveys, offer valuable medical insights into wildfire-related health outcomes; however, they often suffer from limitations such as small sample sizes, limited scalability, high costs, and delayed data availability \cite{garcia2024wildfires}. With the California fire season approaching, there is an urgent need for rapid, scalable methods to detect public health signals and uncover unmet needs in affected communities in real time.

Social media platforms serve as critical sources of information during disasters, significantly supporting disaster response through the rapid dissemination of real-time updates \cite{ye2020social}. Given the sheer volume of data generated on social media platforms far beyond the capabilities of manual examination, existing research leverages machine learning techniques, such as keyword-based methods or classification models, which utilize textual and metadata features, to identify and categorize relevant social media content \cite{vongkusolkit2021situational, huang2015geographic}. Most recently, pretrained large language models (LLMs) have become the state-of-the-art method in natural language processing (NLP). Built on top of these pretrained embeddings, supervised and unsupervised methods are applied as post-processing strategies. Supervised techniques still require a certain quantity of annotated data for fine-tuning, which can be time-consuming in fast-moving emergencies. By contrast, unsupervised methods, such as topic modeling, can quickly identify latent topics based on semantic clustering without annotation efforts \cite{zhou2023guided}. Furthermore, most previous research on disaster events has considerable time lags, with findings published years after the crisis. These findings can inform long-term mitigation, but often lack value for immediate response.

In order to quickly identify and extract actionable information for disaster response and management, a topic classification schema or a coding schema is needed to guide the categorization of social media datasets \cite{vongkusolkit2021situational, verma2011natural, hughes2009twitter}. These guidance schema often include categories such as casualties, damage reports, donation requests, or public emotions. However, designing an effective classification schema and corresponding information extraction methods is complex and highly context-dependent, influenced by factors such as (1) the type of disaster event, (2) the specific goals of the study or analysis, and (3) the characteristics of the social media platform used \cite{vongkusolkit2021situational}.

To address these gaps, this study analyzes Reddit discourse related to the 2025 California wildfires to better understand how affected communities perceived and communicated public health risks in real time. We collect and present the first publicly available Reddit dataset focused on this event as part of a rapid response effort. In addition, this research innovatively designs and employs a hierarchical topic modeling framework, integrating the proposed hierarchical topic modeling framework with latent Dirichlet allocation (LDA) and BERTopic enhanced by LLMs, we track how discussions evolved over the 24 days leading up to full containment of the fires. To sum up, the major contributions of this work include: 

\begin{enumerate}
\item In accordance with the nature of Reddit data, we proposed a dual-topic modeling strategy and developed a multi-hierarchical, multi-label framework that enables timely and effective analysis of social media content. This framework is also adaptable to other unstructured social media data.

\item Using this novel framework, we generated and publicly released multiple labeled datasets of Reddit posts and comments related to the 2025 Los Angeles wildfires, providing a valuable resource for social media research to support disaster response efforts.

\item The scalable framework and resulting dataset for detecting public health–related topics from community-generated social media content is applicable not only to future research on the 2025 Los Angeles wildfires, but also to broader interdisciplinary disaster studies in public health, environmental and exposure science.
\end{enumerate}

\section{Literature Review}\label{sec:review}

Topic modeling is an unsupervised NLP technique that enables the identification and organization of latent thematic structures within large text corpora. This approach is particularly valuable for managing the exponential growth of digital textual data—from emails and reports to social media posts—by converting unstructured content into structured, analyzable representations \cite{vayansky2020review}.

\subsection{Topic Modeling Algorithms}

The core of many topic modeling methods is the “bag of words” paradigm, where the focus is on word frequencies rather than their order \cite{zhang2010understanding}. This paradigm generates two critical matrices: a word-topic matrix that encapsulates topics through their associated words, and a document-topic matrix that reflects each document's composition in terms of these topics. Such representations help extract important insights from vast corpora, especially in the absence of explicit labeling \cite{barde2017overview}.

Historically, topic modeling has evolved significantly—from early models such as Latent Semantic Analysis (LSA) \cite{deerwester1990indexing} and its probabilistic extension, Probabilistic LSA (pLSA) \cite{hofmann1999probabilistic}, to more sophisticated Bayesian approaches like the Gibbs sampling algorithm for the Dirichlet multinomial mixture (GSDMM) \cite{yin2014dirichlet} and Latent Dirichlet Allocation (LDA). LDA, for instance, leverages Dirichlet distributions to iteratively refine the assignment of words to topics based on the probability of topic occurrence in documents and word occurrence within topics \cite{blei2003latent}. Concurrently, Non-Negative Matrix Factorization (NMF) emerged as an alternative linear algebra-based method that decomposes high-dimensional text data into interpretable, lower-dimensional representations \cite{lee1999learning}. 

In recent years, the advent of transformer-based models has ushered in a new era of topic modeling with algorithms such as Top2Vec \cite{angelov2020top2vec} and BERTopic \cite{grootendorst2022bertopic}. Top2Vec detects topics with assumption that each document is primarily centered on a single topic. In contrast, BERTopic performs powerful and flexible modular structure by integration of sophisticated sentence embedding techniques, dimensionality reduction, and clustering algorithms for modern text analytics.

Overall, the evolution from LSA and pLSA to GSDMM, LDA, NMF, and the transformer-based methods illustrates the dynamic development of topic modeling methodologies. Each algorithm has its own merits for applications depending on the trade-offs among scalability, interpretability, and contextual sensitivity. For example, LDA is well-suited for longer texts but is typically efficient only when applied to a relatively small number of documents. In contrast, BERTopic performs better on shorter texts but requires a large corpus to effectively capture and cluster meaningful topics \cite{egger2022topic}.

\subsection{Topic Modeling Applications in Wildfires}

\subsubsection{Dataset Sources}
Wildfire-related topic modeling studies collect dataset primarily from two types of sources: (1) structured reports from official agencies and (2) unstructured, crowdsourced content from social media platforms.

Structured data sources, such as NASA’s Aviation Safety Reporting System (ASRS) and the ICS-209-PLUS incident database, have been used in a limited number of studies focusing on operational safety and technical failures in wildfire contexts \cite{mbaye2023bert, andrade2023machine}). While these datasets are valuable in terms of infrastructure performance and risk reporting, they are less suited for examining public sentiment, communication, and response in real time.

In contrast, most studies rely on social media data, particularly from Twitter, to analyze wildfire trends, public discourse, and crisis communication strategies. These datasets vary in size, ranging from thousands to millions of posts, and span from short-term event monitoring to multi-year longitudinal analyses \cite{zander2023trends, garcia2024wildfires, souza2024using, madichetty2021multi, madichetty2023roberta}. For instance, García et al. (2024) focused on mental health during the 2017 Tubbs Fire in California, while Zander et al. (2023) examined broader situational awareness during Australia’s “Black Summer” bushfires in 2019 and 2020. 

Due to limitations associated with Twitter data—such as character constraints, low information density, credibility issues, and recent restrictions on data access—researchers have begun to explore alternative platforms. Reddit, with its longer-form user content, community-based moderation, and more accessible data policies, presents a compelling alternative for analyzing public discourse around wildfires \cite{lever2022social}.

Collectively, these studies highlight the potential of crowdsourced data to capture real-time public discourse and sentiment, underscoring its value for monitoring and responding to wildfire crises.

\subsubsection{Topic Classification Schema}

After latent topic clusters are generated, it is common to organize them into interpretable categories using domain annotations. Many studies implicitly follow two analytical rationales: situational awareness (SA) and emotional response. SA is defined as “all knowledge that is accessible and can be integrated into a coherent picture, when required, to assess and cope with a situation” \cite{sarter2017situation}, derived from three-level model of perception, comprehension, and projection \cite{endsley1995measurement}. In the context of disaster response, SA captures real-time information such as fire progression, evacuation orders, and resource availability. While SA-based categories are prevalent across studies, shown in Table \ref{tab:sa_review}, they tend to vary widely ranging from three to fifteen categories depending on research goals. This inconsistency reflects a lack of unified structure for organizing real-time information in disaster response.

To capture emotional response, many studies rely on sentiment analysis \cite{lever2022social, garcia2024wildfires, zander2023trends}, which classifies corpus categorizing text into positive, negative, or neutral categories, as well as fundamental emotion such as fear, anger, or sadness. While sentiment analysis provides general emotional trends, it is often insufficient to summarize intricate expressions such as narratives of blame, resilience, or justice \cite{ruiz2017twitter}. Moreover, sentiment labels may be inaccurate in informal, sarcastic, or ambiguous media messages \cite{wankhade2022survey}.

\begin{table}[ht]
\centering
\caption{Summary of wildfire-related studies using Topic Modeling with SA-based rational}\label{tab:sa_review}
\renewcommand{\arraystretch}{1.2} 
\begin{tabular}{p{2.5cm} p{3cm} l l c l}  
\toprule
Author & Disaster & Source & Time Range & Number& Topic Models\\
\midrule
\cite{madichetty2021multi} & Califonia wildfires & Twitter & 2 weeks & 15 & Fine-tuned BERT \\
\cite{lever2022social} & California Wildfire Season (2016) & Twitter, Reddit & 4 months & 5 & LDA \\
\cite{ma2024investigating} & Western Wildfire Season(2020) & Twitter & 1 month & 6 & BERTopic\\
\cite{zander2023trends} & Black Summer(2019-2020) & Twitter & 6 months & 8 & GSDMM \\
\cite{souza2024using} & Wildfires in Brazil National Park & Twitter & 11 years & 6 & BERTopic \\
\cite{garcia2024wildfires} & Tubbs fire & Twitter & 3 weeks & 8 & LDA \\
\cite{andrade2023machine} & Wildfire incidents & ICS-209-PLUS & 8 years & 3 & BERTopic \\
\cite{mbaye2023bert} & Wildfire reports & UAS & 4 years & 7 & BERTopic \\

\bottomrule
\end{tabular}
\end{table}

To address these limitations, we draw on the concept of Crisis Narratives from crisis communication research. Crisis Narratives (CN) refer to “the stories told about crises that carry meaning, encode lessons, and frame larger public and societal understanding of risks, warnings, and potential harm” \cite{seeger2016narratives, liu2020telling}. Common narrative themes, such as blame, victimhood, heroism, and renewal, have been widely used to examine institutional rhetoric and public interpretation in disaster contexts, as shown in table \ref{tab:cn_review}. Despite their relevance, this framework has not yet been systematically applied to social media data in the context of wildfires. In this study, we adopt and extend CN theory to move beyond surface-level sentiment analysis, offering a more nuanced and structured lens for interpreting emotionally charged public discourse during crises.

\begin{table}[ht]
\centering
\caption{Summary of Crisis Narrative Studies}\label{tab:cn_review}
\renewcommand{\arraystretch}{1.2} 
\begin{tabular}{p{2cm} l p{5.5cm} p{6cm}}  
\toprule
Author & Event & Data Source & Category \\
\midrule

\cite{wolfe2016organizing} & Disaster & Interview, participant observation, public meetings, media reports and organizational documents & Blame and Accountability, Victim-Centered Disaster, Renewal and Growth \\
\cite{liu2020telling} & Pandemic & Quantitative survey & Blame, Renewal, Victim, Hero, Memorial \\
\cite{an2024dynamics} & Pandemic & Twitter & Blame, Renewal, Victim, Hero, Memorial \\

\bottomrule
\end{tabular}
\end{table}

To our knowledge, no existing study has jointly formalized SA and CN into a unified category framework. This gap limits our ability to holistically analyze how individuals both perceive and emotionally respond to crisis events in real time. Our study fills this gap by developing a multiple-layer topic modeling framework that explicitly integrates both SA and CN dimensions. This framework supports multi-label annotation, human-in-the-loop validation, and scalable modeling for large-scale social media discourse. We elaborate on its design and implementation in the following Methods section.

\section{Methods}\label{sec:headings}

\begin{figure}[ht]
\centering
\includegraphics[width=1\textwidth]{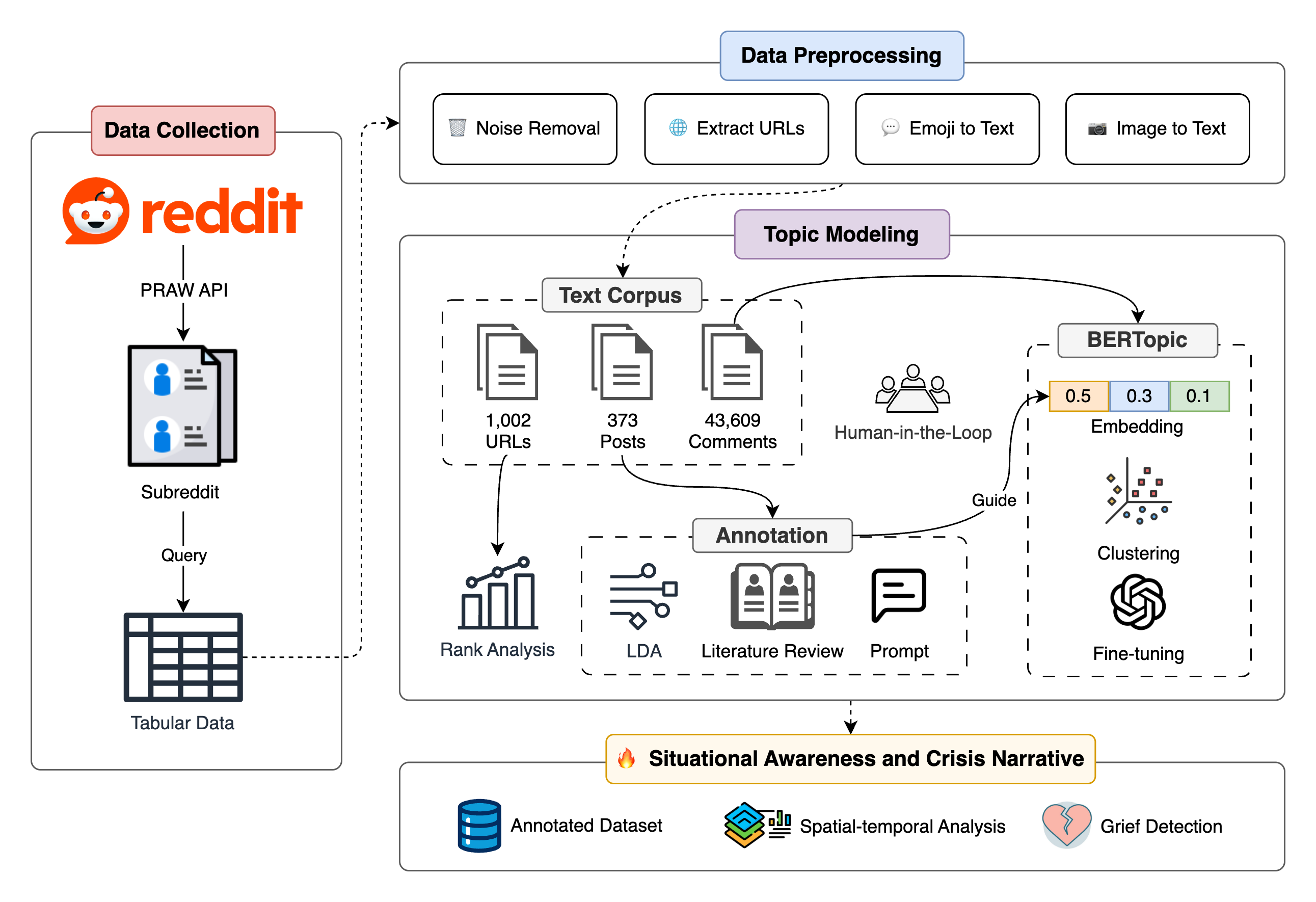}
\caption{A Topic Modeling framework enhanced by Large Language Models and Human-in-the-Loop annotation}\label{fig_framework}
\end{figure}

Our framework, shown in Figure~\ref{fig_framework} is mainly comprised of four components: data collection, data pre-processing, topic modeling and topic analysis.

\subsection{Data Collection}
We collected social media data from Reddit using the Python Reddit API Wrapper (PRAW). A set of predefined keyword-based queries was applied to selected subreddit communities relevant to the 2025 Los Angeles wildfires. The data collection period ranged from January 1 to February 7, encompassing one week prior to and one week following the peak fire events. Both original posts and their associated comments were retrieved, resulting in a corpus of 385 posts and 114,879 comments. Tabular metadata, including user-level statistics (e.g., user ID, comment counts, timestamps, subreddit ID), was also collected to support quantitative spatial-temporal analysis.

\subsection{Data Preprocessing}
We adapted a multi-step preprocessing to clean the Reddit data for analysis. In the first step, we removed special characters, HTML tags, and Markdown symbols. We excluded instances with fewer than 10 words based on the recommended effective semantic length (15-20 words) \cite{markel2009technical}. Unlike traditional NLP methods that completely delete emojis and URLs, we extracted and saved 1,002 unique URLs embedded in posts and comments for rank analysis. Emojis were converted to their corresponding textual descriptions in order to preserve emotional signals. Notably, for image-based posts, we applied image-to-text conversion supplemented by human interpretation and prompt-based reasoning. These steps ensured that the final textual corpus was if excellent quality, semantically intact, and suitable for subsequent topic modeling and analysis.

\subsection{Topic Modeling}

Since Reddit comments are typically generated in response to the original posts and closely follow their discussion threads, we assume that the topics of the posts also represent the primary thematic context for their associated comments. Given the structural and linguistic differences between posts and comments, we adopted a dual-topic modeling strategy for them separately.

\subsubsection{Post-level Modeling}
On the post level, given the limited number of posts (373) but their relatively longer length, we first used Latent Dirichlet Allocation (LDA) model from the Gensim library to extract latent topics. We tuned \texttt{number\_topics} in the range of 5 to 30. The optimal number was 9 based on coherence scores (maximum of 0.49) and interpretability. The coherence value for a single topic measures how the top scoring words in this topic are
semantically similar to each other, with higher coherence indicating that the topic is more interpretable and meaningful for human understanding \cite{zhou2023guided}.

Next, we integrated human-in-the-loop (HITL) to review top-ranked keywords and representative posts. We found that, with an average length of 120 words, the posts often include rich, in-depth and complex emotional content. Drawing on categories established by previous studies, we decided to apply multiple-label category to the post-level data, and developed a multi-layer hierarchical classification system with two main categories: Situational Awareness (SA) and Crisis Narratives (CN). Table~\ref{tab:category} presents the three-layer schema.

Guided by HITL, we further identified six subcategories under SA: fire operations, public health and safety, emergency resources, recovery, loss and damage, and influential figures. The original definition of CN includes five narrative forms: blame, victim, hero, renewal, and memorial. However, we did not observe the memorial narrative in our dataset, perhaps as a result of the recent crises.

To better capture emotional sensitivity and health-related concerns beyond traditional narrative types, we also introduced two supplementary binary labels: grief signals and potential mental health risks. These were annotated based on explicit expressions of emotional loss, mourning, psychological distress, or self-identified vulnerability \cite{cunsolo2018ecological}.

This multi-layer, multi-label framework was refined collaboratively but independently by three authors from different fields of study. To ensure consistency in annotation, we developed a decision tree–based guideline to assist the labeling process. The detailed annotation flowchart is provided in the Supplementary Information \ref{d_tree}. It not only guided the annotation of posts but also provided representative seed terms for comment-level clustering.

\begin{table}[ht]
\centering
\begin{threeparttable}
\caption{Hierarchical category framework of Reddit wildfire discourse}
\label{tab:category}
\begin{tabular}{lll}
\toprule
\textbf{First Level} & \textbf{Second Level \tnote{1}} & \textbf{Selected Keywords} \\
\midrule
\multirow{6}{*}{Situational Awareness} 
& Fire operations & watchduty, calfire, contaiment \\
& Public health and safety & air quality, smoke, medical\\
& Emergency resources & eggs, hydrant, laundry\\
& Recovery & insurance, donation, restore\\
& Loss and damage & burned down, cars, trails \\
& Influential figures & influencer, celebrity, trump \\
\midrule
\multirow{4}{*}{Crisis Narrative} 
& Blame & mayor, edison, drone \\
& Renewal & clean, therapy, relief\\
& Victim & lost, damage, structure\\
& Hero & inmate, firefighters, volunteer \\
\bottomrule
\end{tabular}
\begin{tablenotes}
    \item [1] The detailed definition is provided in the Supplementary Information \ref{d_tree}
\end{tablenotes}
\end{threeparttable}
\end{table}

\subsubsection{Comment-level Modeling}
We employed BERTopic topic modeling with guided approach to explore latent topics. Our pipeline consisted of five modular steps, each carefully selected and tuned to balance semantic coherence and interpretability.

\textbf{Step 1: Sentence embedding.} We utilized the \texttt{all-mpnet-base-v2} model from the SentenceTransformers library to encode each comment into dense vector representations. This general-purpose model demonstrated the best performance at sentence level. \cite{ashqar2023comparative}.

\textbf{Step 2: Dimensionality reduction.} We selected Uniform Manifold Approximation and Projection (UMAP) to reduce embedding dimensionality while preserving the global structure of the semantic space. We tuned two important hyperparameter \texttt{n\_neighbors} and \texttt{min\_dist}. The \texttt{n\_neighbors} determines the number of neighbors whose local topology is preserved, with higher values placing greater emphasis on global structure. Meanwhile, \texttt{min\_dist} influences the minimum distance between samples in the embedded space, thereby affecting the compactness and spread of topic clusters.

\textbf{Step 3: Clustering.} Clusters of semantically similar comments were identified using HDBSCAN, a density-based clustering algorithm. We tuned the \texttt{min\_cluster\_size} and \texttt{min\_samples} to accommodate varying topic densities.

\textbf{Step 4: Weighting.} 
We followed best practices by using a \texttt{CountVectorizer} to capture both unigram and bi-gram patterns, combined with the \texttt{ClassTfidfTransformer} (c-TF-IDF) to compute distinctive keywords for each topic to  get an accurate representation.

\textbf{Step 5: Topic representation.} To further enhance topic interpretability, we used OpenAI’s \texttt{gpt-4o-mini} model for label generation and summarization. Structured prompts were used to generate concise topic names and human-readable descriptions, based on top-ranked keywords and representative documents. Maximal Marginal Relevance (MMR) was applied to diversify selected keywords.

In steps 2 and 3, we tuned hyperparameter, shown in table \ref{tab:hyperparam-search} and supplementary figure \ref{sf_parameter}, across \texttt{n\_neighbors} $\in$ \{15, 20, 25, 30\}, \texttt{min\_dist} $\in$ \{0.0, 0.01\}, and \texttt{min\_cluster\_size} $\in$ \{50, 100, 150, 200, 250, 300, 350, 400\}, while setting \texttt{min\_samples} to half of the cluster size. For each configuration, we trained a separate BERTopic model and evaluated its performance using the best-performing coherence metric $C_v$ \cite{roder2015exploring} with the \texttt{CoherenceModel} from the Gensim library. The final model, configured with \texttt{n\_neighbors = 30}, \texttt{min\_dist = 0.01}, and \texttt{min\_cluster\_size = 300}, achieved the best balance between topic coherence (0.628), interpretability, and semantic diversity. The optimal
number of latent topics is 30, we further mapped the 30 latent topics to the pre-defined hierarchical category forwork, the mapping examples are shown in Supplementary Information \ref{latent_mapping}.

\begin{table}[ht]
\centering
\caption{Search space for BERTopic hyperparameter tuning.}
\label{tab:hyperparam-search}
\begin{tabular}{llll}
\toprule
\textbf{Component} & \textbf{Parameter}         & \textbf{Values Explored} & \textbf{Optimal}  \\
\midrule
\multirow{2}{*}{UMAP} & \texttt{n\_neighbors}     & 15, 20, 25, 30  & 30\\
                      & \texttt{min\_dist}        & 0.0, 0.01 & 0.01\\ 
\midrule
HDBSCAN               & \texttt{min\_cluster\_size} & 50, 100, 150, 200, 250, 300, 350, 400 & 300\\
\bottomrule
\end{tabular}
\end{table}

\subsection{Topic Analysis} 
To reveal the frequencies, interactions, and temporal changes of different topics discussed throughout the progression of various fire events, we assumed equal weighting for posts and comments and combined them into a single dataset. A range of statistical and visualization methods was employed, including UpSet plots, histograms, and BERTopic visualizations across time and locations. In particular, we focused on the public health and safety category, which frequently intersects with other categories in the multi-labeled dataset.

\section{Results}\label{sec4}

\subsection{Hierarchy Topic Modeling Analysis}

The final Reddit dataset includes 43,982 records of posts and comments contributed by 43,911 unique users. The effective retention rates are 373 out of 385 posts (96.88\%) and 43,609 out of 114,879 comments (37.96\%), with an additional 2,466 comments marked as deleted. All retained data were annotated for situational awareness (SA). A subset of this dataset (40,688) was annotated for crisis narrative (CN), comprising 228 posts and 40,460 comments. In this subset, 26,257 instances were annotated with grief theme, and 19,683 instances were annotated as implying mental health risks.

\begin{figure}[ht]
\centering
\includegraphics[width=1\textwidth]{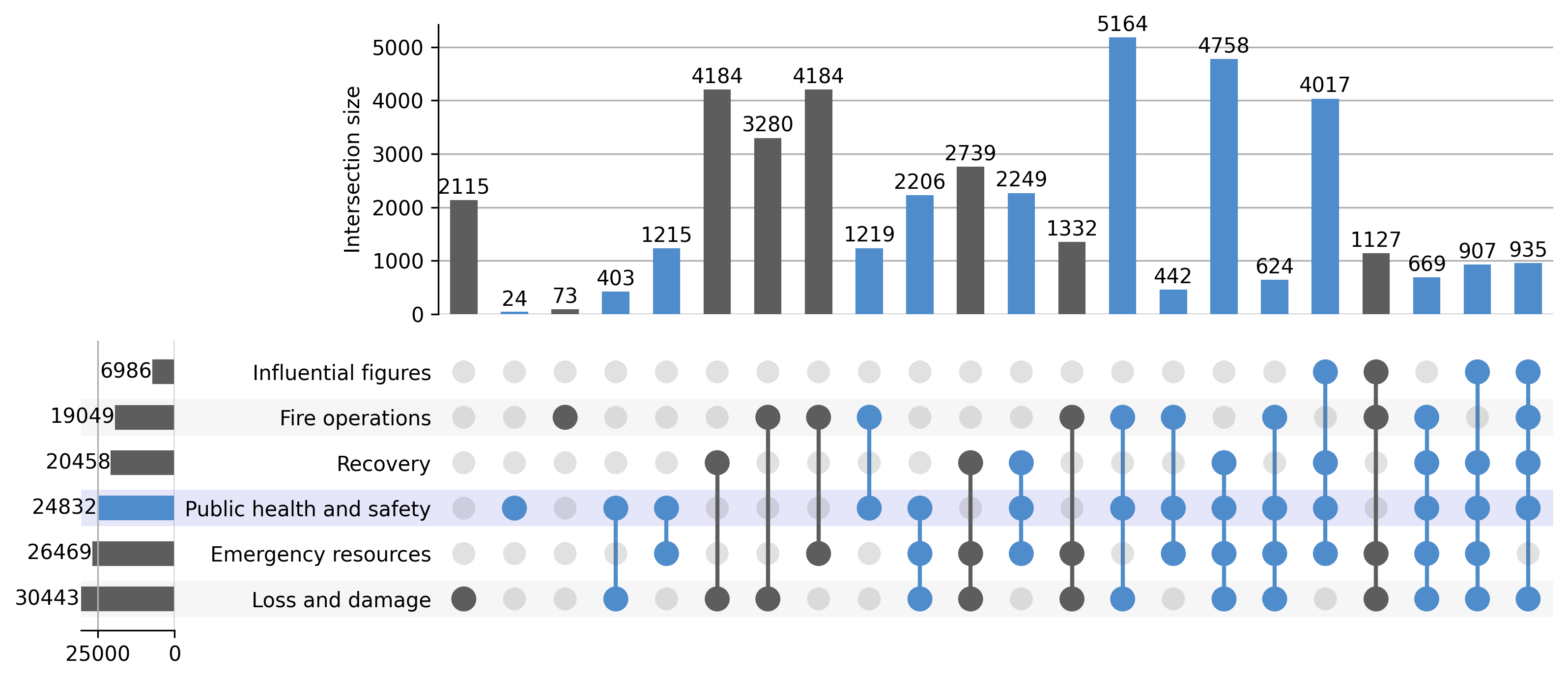}
\caption{\textbf{Situational awareness (SA) intersection patterns revealed by UpSet plot.} The UpSet plot visualizes topic co-occurrences across three components: the left horizontal bar shows the size of each individual set (topic category), the bottom matrix indicates which sets are involved in each intersection, and the top histogram displays the size of these intersections. For example, the set size of Public Health and Safety is 24,832 instances. The combination of Public Health and Safety and Fire Operations (indicated by connected dots in the matrix) accounts for 1,219 instances.}\label{fig_sa}
\end{figure}

Figure \ref{fig_sa} shows that public health and safety is third among six SA categories, accounting for 24,832 instances. It is also the third most engaged topic, with 56.6\% of users contributing to related posts or comments. Although it is not the most discussed category, it has the greatest degree of correlation with other categories. Meanwhile, we also calculated user engagement across the dataset. In total, we identified 43,911 unique users who participated in wildfire-related discussions. Fire operations, public health and safety, and loss and damage make up the largest intersection set, with 5,164 instances. The second largest intersection had 4,758 instances involving recovery, public health and safety, and loss and damage. Additionally, intersections involving subsets of public health and safety, emergency resources, and loss and damage comprise 9,164 records. Overall, public health and safety appears in five of the top ten most frequent intersections. Moreover, while public health and safety rarely appears as a standalone category (only 24 instances). These findings suggest that public health topics are consistently intertwined with other SA categories and highlight the integrative nature of public health within wildfire communication and discourse during wildfire crises.

\begin{figure}[ht]
\centering
\includegraphics[width=.7\textwidth]{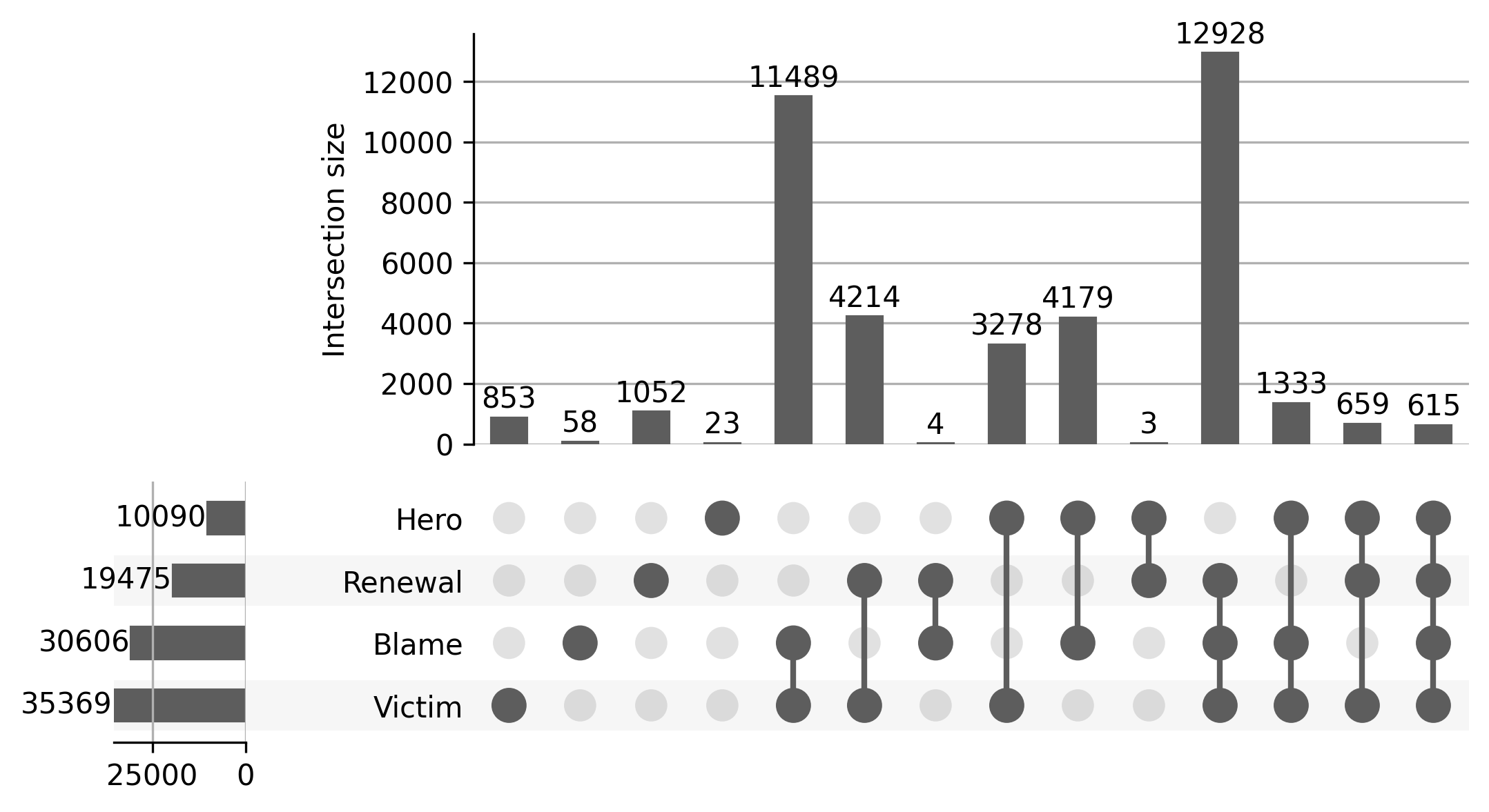}
\caption{\textbf{Crisis narrative (CN) intersection patterns revealed by UpSet plot.} The UpSet plot visualizes topic co-occurrences across three components: the left horizontal bar shows the size of each individual set (topic category), the bottom matrix indicates which sets are involved in each intersection, and the top histogram displays the size of these intersections. For example, the set size of Victim is 35,369 instances. The combination of Victim and Blame (indicated by connected dots in the matrix) accounts for 11,489 instances.}\label{fig_cn}
\end{figure}

In Figure \ref{fig_cn}, four crisis narrative categories are presented in descending order of frequency: victim (35,369), blame (30,606), renewal (19,475), and hero (10,090). The largest intersection set consists of the simultaneous occurrence of renewal, blame, and victim narratives (12,928 instances), while the second-largest intersection comprises the combination of blame and victim narratives (11,489 instances). The combined total of these two largest intersections exceeds the sum of all other narrative combinations, highlighting a strong linkage among victimhood, blame attribution, and renewal attitudes. This distribution illustrates the interwoven emotional framings within social media storytelling during the wildfire events.

\subsection{Public Health Latent Topics in Situational Awareness}

\begin{figure}[ht]
\centering
\includegraphics[width=1\textwidth]{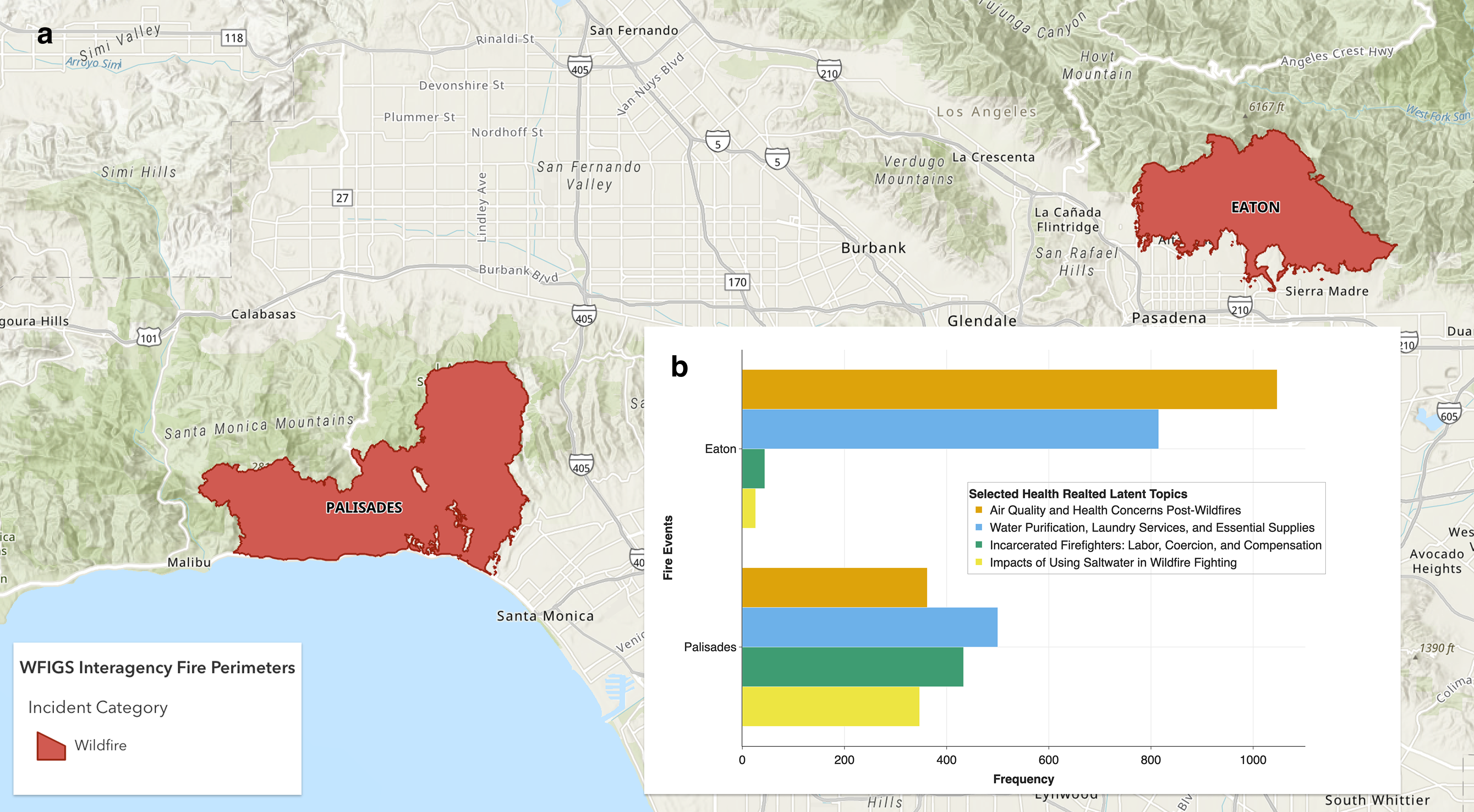}
\caption{\textbf{Spatial pattern}: a. Geographic extent and burned area comparison for the Palisades and Eaton Fires, mapped as of January 31; b. Comparison of public health realted topic distributions between the Palisades and Eaton Fires}\label{fig_map}
\end{figure}

\begin{figure}[ht]
\centering
\includegraphics[width=1\textwidth]{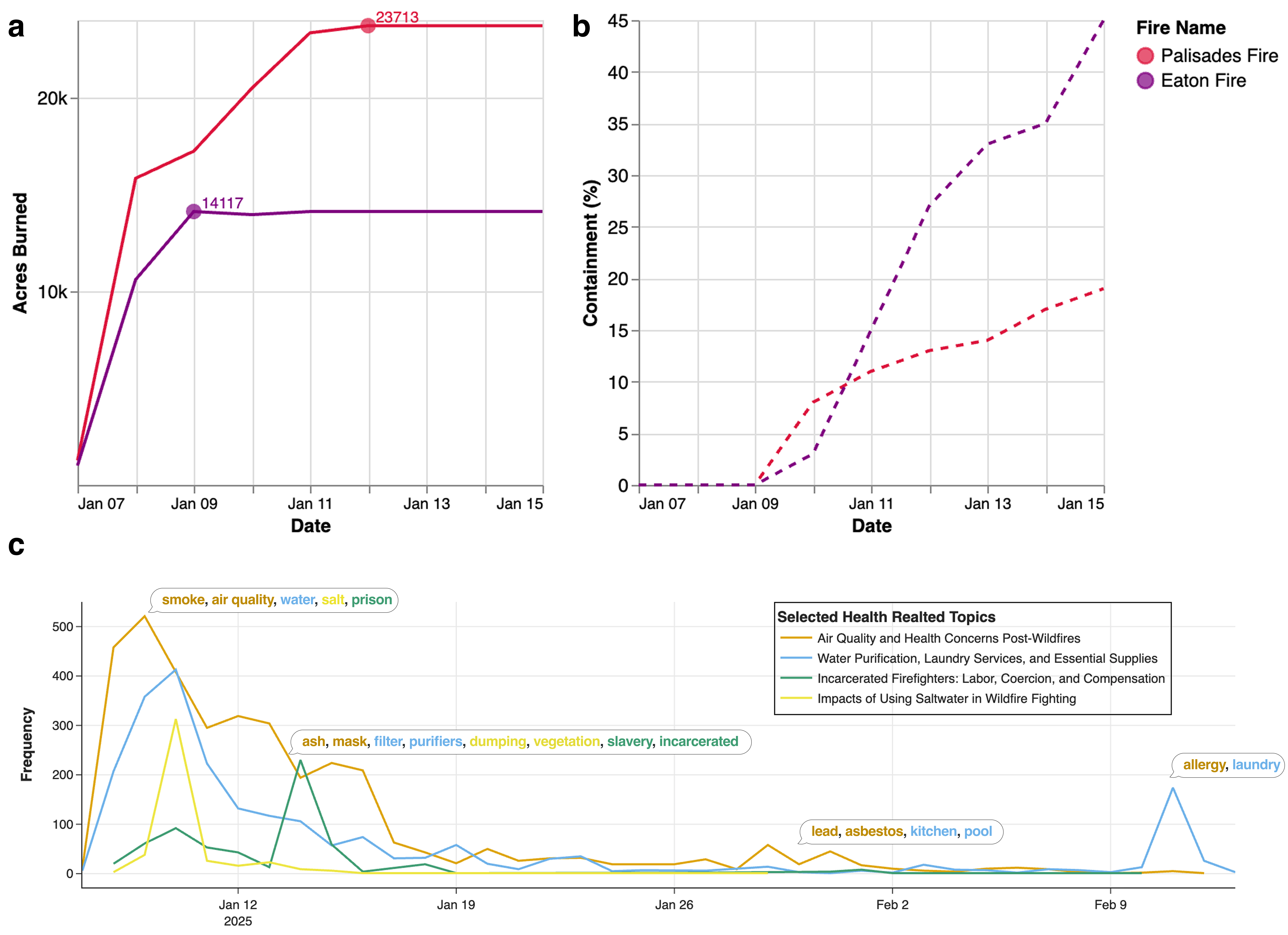}
\caption{\textbf{Temporal pattern}: a. Summary of progression and b. Suppression details for the Palisades and Eaton Fires, as inferred from Reddit posts c. Temporal trends and domain keywords of selected public health-related topics}\label{fig_time}
\end{figure}
 
Figures \ref{fig_map}a and \ref{fig_time}a and \ref{fig_time}b show the geographic locations of the Eaton and Palisades fires, their respective burned areas and containment statuses between January 7 and January 15. The Palisades fire expanded faster and more extensively, reaching 23,713 acres on January 11 and remaining at that size until January 15. In contrast, the Eaton fire grew and ceased at 14,117 acres by January 9. By January 15, the Eaton Fire reached 45\% containment, while the Palisades Fire was only at 19\%.

We examined Reddit data collected from geographically relevant subreddit communities—r/altadena and r/pasadena for the Eaton Fire, and r/PacificPalisades for the Palisades Fire. A total of 28,648 instances discussed both fires. In addition, 17,210 instances focus exclusively on the Eaton Fire, while 26,038 instances are specific to the Palisades Fire. To better understand distinctions in public discourse between these two major fire events, we excluded both the overlapping 28,648 instances and the 5,298 instances associated with other fires, such as the Hughes Fire. Furthermore, we selected public health related latent topics, such as air quality, water safety for environmental health, firefighters for occupational health, and seawater effects on vegetation for one health. 

Figure \ref{fig_map}b illustrates how public health related discussions varied between the two fire events based on their distinct geographic locations. Discussions of the Eaton Fire, which occurred in a canyon, primarily focused on air pollution and water safety. In contrast, discussions of the Palisades Fire, located in a coastal area, exhibited a more balanced distribution of topics. Some timely unique topics emerged in the Palisades Fire discussions, including issues related to saltwater contamination and incarcerated firefighters.

Figure \ref{fig_time}c illustrates the temporal variation in public health discussions. Both fire events exhibited similar trends over the month, with concerns about air quality emerging rapidly and peaking early in the timeline, followed by discussions on water safety. In a whole, the volume of these discussions closely aligned with the acres burned during the initial days of the fires. Although the overarching topics remained consistent, the representative keywords evolved at different stages of the events. For topics on air quality, discussions progressed from smoke to masks, then to indoor hazards such as lead and asbestos, and finally to allergies. Health concerns over water shifted from drinking water to kitchen use, swimming pools, and ultimately laundry services. Discussions on seawater contamination and incarcerated firefighters also appeared rapidly at the onset of the fires but dissipated just as quickly after the fire reached their maximum extent.

\subsection{Grief and Mental Health in Crisis Narrative}

\begin{figure}[ht]
\centering
\includegraphics[width=1\textwidth]{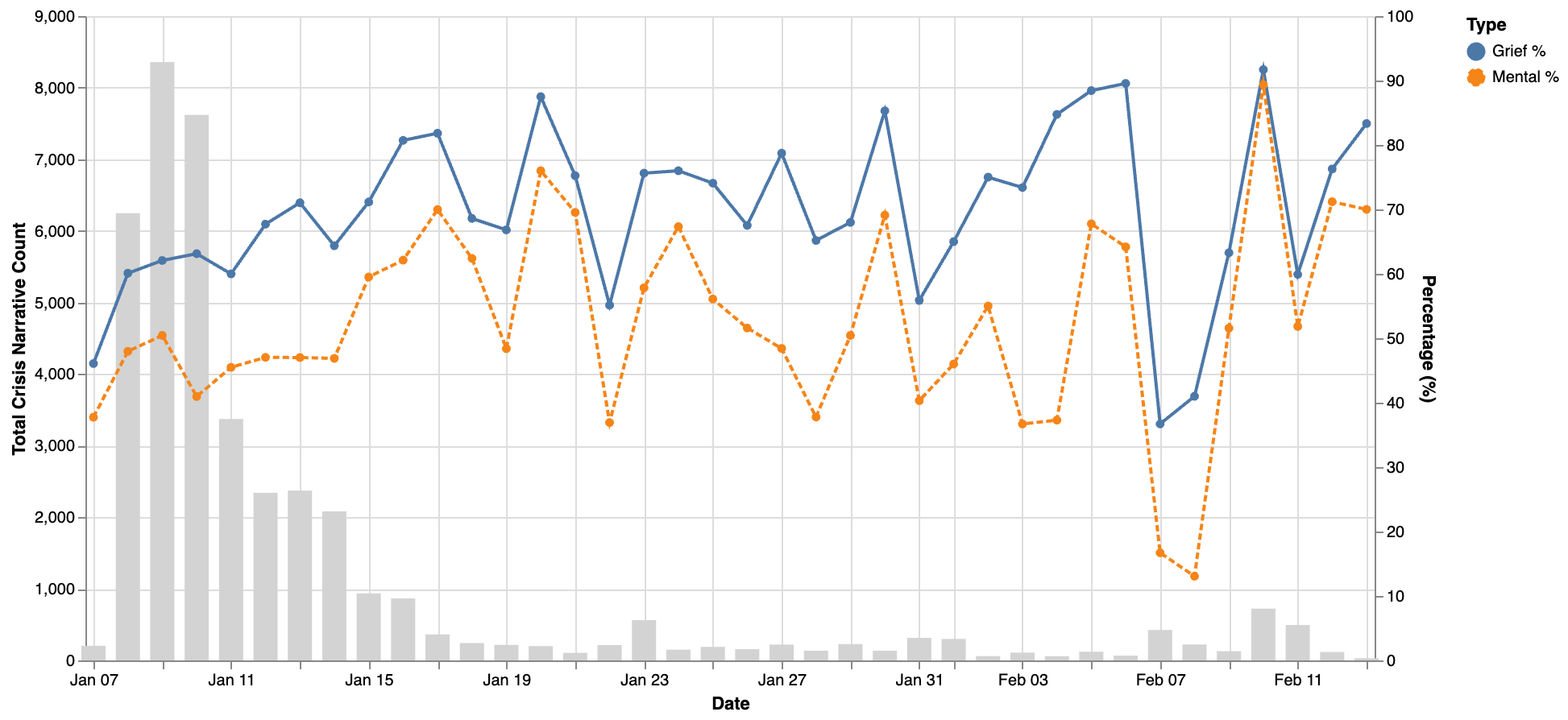}
\caption{Temporal trends of Crisis Narrative, grief percentage and mental health risk percentage}\label{fig_day}
\end{figure}

As shown in Figure \ref{fig_day}, The total amount of crisis narrative content follows the wildfire pregression,  initial days of the wildfire events, particularly between January 8 and January 10, with daily counts rang from 6,000 to 9,000. After January 11,  daily counts gradually declined and stabilized at lower levels through February.

However, the proportions of grief and mental health risks in the content increase gradually before mid-January and then fluctuate dramatically. The content containing grief (blue line) was consistently high, frequently stay above 60\%  after the second day of the fire in January and reaching 90\% once in early February. Mental health risk content (orange line) followed a similar pattern, but with more fluctuation. Notably, both grieving and mental health percentages dropped significantly around February 7, but rebounding immediately afterward.
 
\begin{figure}[ht]
\centering
\includegraphics[width=.6\textwidth]{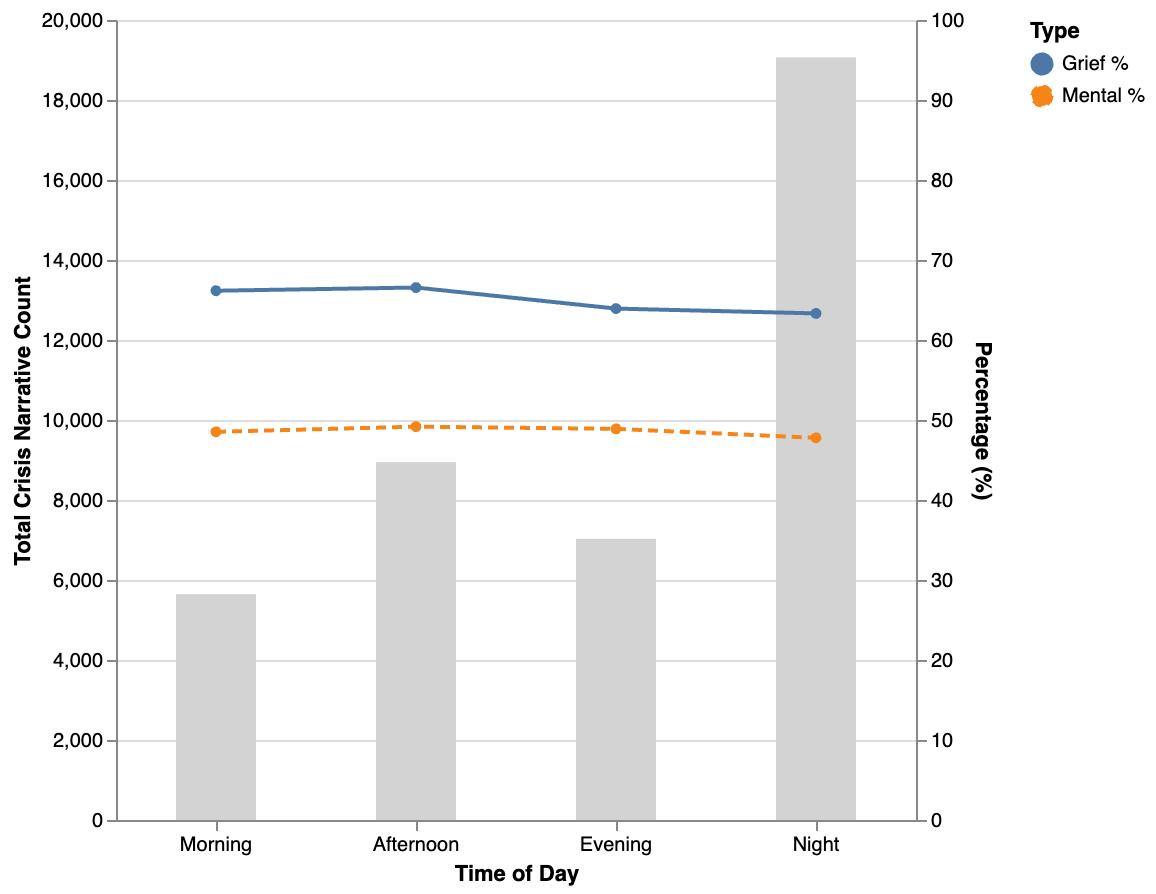}
\caption{Distribution of crisis Narratives, grief, and mental health risk percentages by time of day}\label{fig_dt}
\end{figure}

Figure \ref{fig_dt} shows the temporal distribution of grief-related and mental health risk content in four time segments: morning, afternoon, evening, and night.  The discussion mainly occurs at night with nearly 20,000 instances, almost twice as many as any other time of day. Despite the volume differences, the proportions of grief-related and mental health risks content are relatively stable throughout the day, averaging approximately 65\% and 50\%, respectively.

\subsection{Representative Narrative Snapshots}
\begin{figure}[ht]
\centering
\includegraphics[width=1\textwidth]{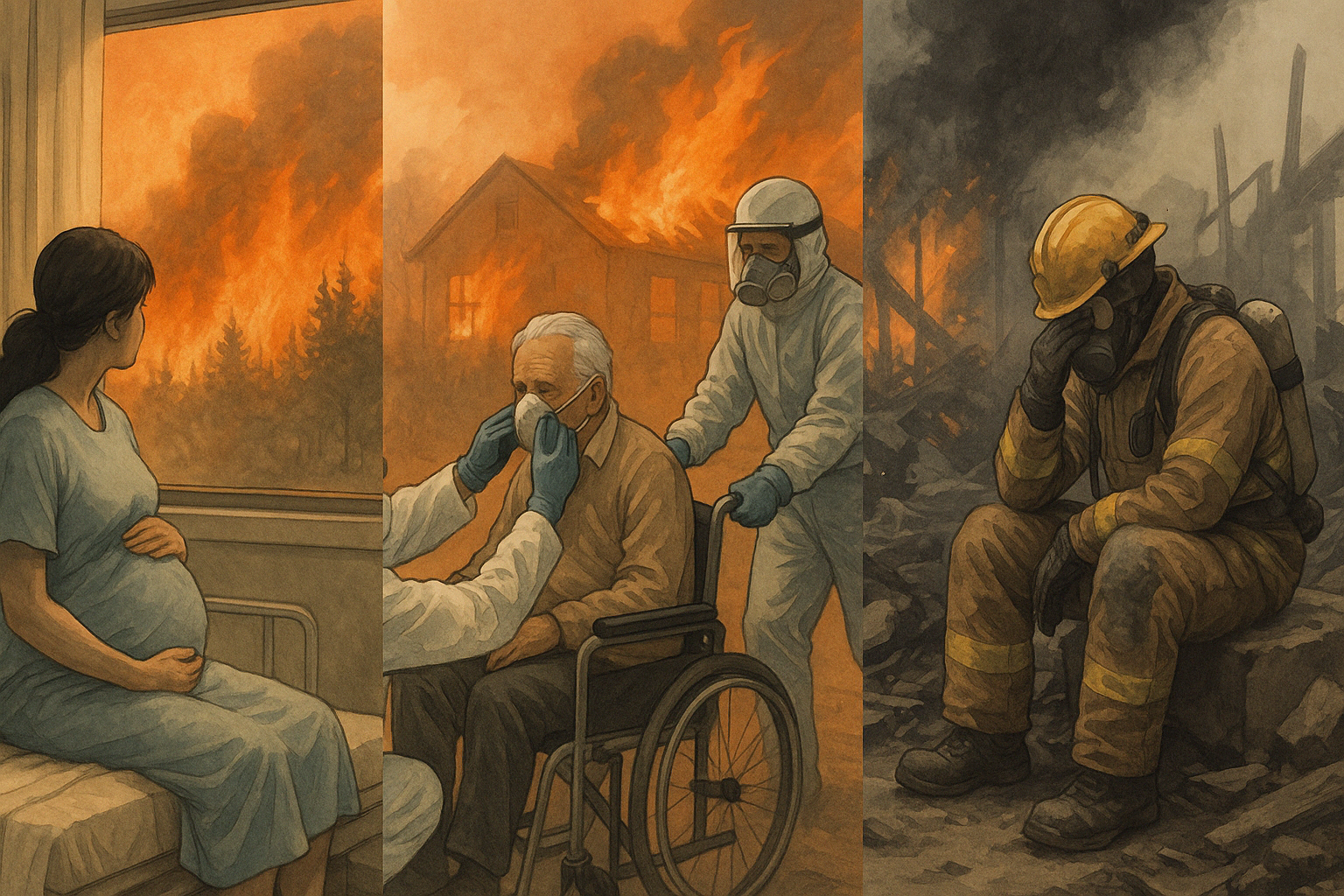}
\caption{Illustrative depictions of three vulnerable groups identified in Reddit discourse during the Wildfires: pregnant women, elderly individuals, and exhausted first responders (generated by GPT-based Model)}\label{fig_tale}
\end{figure}

In this section, we present selected representative anonymized posts and comments. As shown in Table \ref{tab:example_topics}, these documents reflect widespread public concern regarding both the short- and long-term environmental and occupational health impacts of the wildfires. Air quality, asbestos exposure, and chemical pollutants are widespread topics for affected community, particularly among vulnerable populations such as pregnant women, new moms and newborns, the elderly, and people with pre-existing diseases like asthma. Water contamination from debris also posed risks to municipal drinking water systems. In addition, people share evidence of respiratory symptoms and adverse health effect from first responders, such as journalists and firefighters. Last but not least, discussions around the emergency use of seawater for firefighting raised public concern about long-term ecological impacts, especially the potential harm to the growth of inland vegetation. In order to better empathize with vulnerable populations while also protecting data privacy during this crisis, we generated three illustrative scenes, shown in Figure \ref{fig_tale}. From left to right, the images depict high-risk health environments for a pregnant woman, an older adult, and frontline workers, including medical volunteers and firefighters.

\begin{table*}[ht]
\caption{Examples of topics and representative documents}\label{tab:example_topics}
\begin{tabular*}{\textwidth}{@{\extracolsep\fill}l p{12cm}}
\toprule
\textbf{Health Topic} & \textbf{Representative Text} \\
\midrule
Environmental Health & Hey mamas, I'm 8 months pregnant and feeling incredibly lucky that our house is still standing. I don’t think many people fully realize what we’ll be exposed to over the next few years in terms of water quality, air pollution, and other environmental factors.  \\
Environmental Health & Hi all, I’m kind of freaking out about the air quality. I know the AQI has been looking better, but as I understand it, that doesn’t take into account asbestos and other harmful chemicals. With all those old houses and electronics burning, I’m paranoid that I’ll develop cancer in a few years from breathing it in. I live in Pasadena, about 3.5 miles from the Eaton Fire. I also have asthma, and while I haven’t had an episode in many years, it’s still a concern.\\
Environmental Health & Debris and elevated turbidity from the Eaton Fire potentially impacted Pasadena Water and Power’s (PWP) drinking water system in the Eaton Fire evacuated areas.\\
Occupational Health & Some people are in the anger stage of their grieving and asking why there were no firefighters or trucks on their specific street. I think the all of the firefighters and companies out there are doing an amazing job and doing the best they can under the circumstances with multiple large fires, 80+ mph wind gusts, and complex evacuation logistics. \\
Occupational Health & This has been a concern of mine after watching the livestreams tonight, especially that reporter who was struggling to breathe during a live shot. They were there for 8 days, and when he came home, his lungs were shot and he spiked a 105° fever. \\
One Health & there are still impacts and if sea water is the first option the salt will accumulate and then prevent more vegetation from growing. \\
One Health & While seawater can be used in certain situations, it's generally not the preferred option due to these potential long-term environmental impacts. \\
\bottomrule
\end{tabular*}
\end{table*}

\subsection{URLs and External Information Source Analysis}

\begin{table}[ht]
\centering
\caption{Top health-related URLs cited under each Situational Awareness category}
\label{tab:sa_urls}
\begin{tabular}{lll l}
\toprule
\textbf{SA Category} & \textbf{Top 1} & \textbf{Top 2} & \textbf{Top 3} \\
\midrule
Fire Operations & \texttt{watchduty.org} & \texttt{fire.ca.gov} & \texttt{arcgis.com} \\
Public Health and Safety & \texttt{cityofpasadena.net} & \texttt{broadcastify.com} & \texttt{airnow.gov} \\
Emergency Resources & \texttt{google.com} & \texttt{genasys.com} & \texttt{flightradar24.com} \\
Recovery & \texttt{gofundme.com} & \texttt{alertcalifornia.org} & \texttt{lafd.org} \\
Influential Figures & \texttt{youtube.com} & \texttt{x.com} & \texttt{wikipedia.org} \\
Loss and Damage & \texttt{instagram.com} & \texttt{imgur.com} & \texttt{latimes.com} \\
\bottomrule
\end{tabular}
\end{table}

Among the 1,002 extracted URLs, we grouped them into situational awareness (SA) categories based on the services they provide and their intended audiences. As shown in Table~\ref{tab:sa_urls}, both crowdsourced platforms and official institutional sources played complementary roles in information dissemination. Official domains were especially dominant in the Public Health and Safety category, whereas crowdsourcing platforms were more frequently cited in categories such as Emergency Resources, Loss and Damage, and Recovery. Platforms such as YouTube, X (formerly Twitter), Wikipedia, Google, and GoFundMe appear frequently in the dataset, reflecting their roles as popular social media and mutual support websites. Of note, while Watch Duty (watchduty.org) appeared more frequently than any other site in our dataset, it has not been systematically analyzed—or even mentioned—in prior studies. It is a crowdsourced platform created by firefighters, dispatchers, and first responders, who monitor radio scanners and official sources to deliver real-time updates on fire progression, evacuation notices, air quality, and other emergency information. Its multifunctionality and trustworthiness help explain its popularity and high visibility during the wildfire response.

\begin{figure}[ht]
\centering
\includegraphics[width=.6\textwidth]{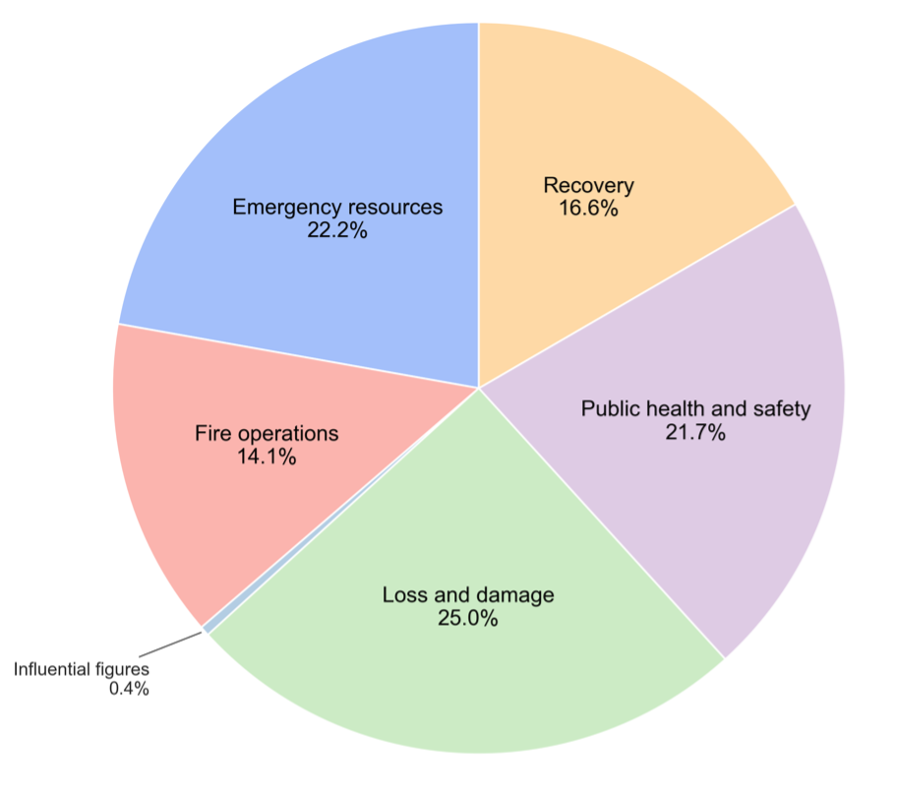}
\caption{Distribution of Health-Related URLs Across Situational Awareness Categories}\label{fig11}
\end{figure}

In Figure \ref{fig11}, health-related websites were mentioned across nearly all SA categories with relatively even distribution—except for Influential Figures, where such content was rarely cited. This again reflects the strong intersection between health concerns, emergency services, and damage-related discussions throughout the fire crisis.

\section{Discussion}\label{sec5}

Our study reveals widespread public health concerns expressed on social media platform, Reddit, in the 2025 Los Angeles widefire crsis. The deletion of comments denotes that the opinions on social media are perishable, which means that they can change and even vanish over the time. This highlights the importance of timely data retrieval and analysis to capture public discourse in real time. Although Reddit posts are generally of higher quality than comments in terms of contextual depth and coherence, they are substantially fewer in number. 

\subsection{Revealing Public Health Concerns}
Back to the effective datasets, approximately half of all identified instances are directly related to public health and safety, intersecting with other SA categories, such as fire operations, emergency resources and loss and damage. Sepecifically, we observe multiple latent topics in the public health and safety category. These include concerns about environmental health, particularly regarding air and water quality; occupational health, with a focus on firefighters, journalists, and emergency responders; and one health, debating on how the use of seawater could nagetively impact inland vegetation. Furthermore, several discussions highlight disparities in access to real-time medical support and protective resources. These findings align with prior research emphasizing the adverse health outcomes of wildfire smoke exposure, particularly among vulnerable populations \cite{rappold2017community, zhang2025respiratory, lei2024wildfire}.

These public health concerns also suggest geographic variation between the two fire events. The Palisades Fire, located near the Pacific Ocean, prompted debates over wheather the use of seawater for emergency suppression can inflict long-term ecological loss. Additionally, the controversial deployment of incarcerated firefighters revealed occupational disparities in regional policy. In contrast, the Eaton Fire, which occurred in a canyon region, generated more frequent discussions about air quality and smoke exposure, largely due to its dense botanical landscape. Despite their geographic differences, both crises revealed similar concerns about safe water supply and respiratory health.

\subsection{Grief and Mental Health Risks Detection}

The crisis narrative category captures the complex emotions and reactions of individuals in such an unexpected natural diaster.  Notably, our temporal analysis suggests an absence of a preparedness phase because there is no instance on Reddit before January 7. In general, there should be 3 stages, preparedness, response and recovery, however, there is no discussion before January 7. The gap between anticipated and actual outcomes can lead to cognitive dissonance \cite{dwivedi2018involvement}, as individuals struggle to reconcile their proactive measures with the devastation. Consequently, learned helplessness may erode confidence in future resilience efforts within affected communities \cite{landry2018learned, ramachandran2021learned}. Similarly, we detect strong signal of grief emotion and mental health risks in the Reddit data. While the physical spread of the fire lasted only a few days, emotional and psychological recovery may take far longer as other primary mental health studies discovered in many wildfire events around the world  \cite{wettstein2024psychotropic, to2021impact}.

\subsection{Topic Classification Schema}

The multi-hierarchical topic classification schema and multi-labeling strategy proved to be effective in understanding situational awareness (SA) and crisis narratives (CN), revealing the many facets of humanity during the wildfire crisis. The first and second level classification categories offer a transferable framework that can be applied to similar wildfire events and other disasters in densely populated areas. Meanwhile, the selected keywords from LDA and latent topics from BERTopic capture the unique characteristics of this specific event, providing deeper insights into localized public concerns and response dynamics. Moreover, the detection of grief and potential mental health risks provides a novel yet cautious approach to stand and empathize with affected communities and better understand their emotional needs. Such insights can help advocate more compassionate public health interventions to support mental well-being in disaster events.

\subsection{Limitations and Future Work}
Our study has several limitations. First, while Reddit provides a rich and near–real-time stream of hyper-local discourse during wildfire events, its user base may not fully represent the demographic, socioeconomic, or geographic diversity of the affected population. The platform tends to overrepresent younger, more technologically literate individuals, potentially excluding samples from older adults, marginalized groups, or those without consistent internet access. As such, our findings may be biased toward certain perspectives and should be interpreted accordingly. Second, the dataset lacks precise geolocation information. Although we mitigated this limitation by leveraging subreddit names for specific places, this proxy does not validate geographic location. Third, while our topic modeling approach was supported by human-in-the-loop verification to enhance interpretability and  robustness, it remains inherently limited by the capabilities of unsupervised machine learning. Subtle semantics, sarcasm, or culturally specific expressions may be under-detected or misclassified. Future research should incorporate complementary data sources, such as X/Twitter, YouTube, or community-level Facebook groups, as well as clinical reports and survey-based public health assessments to triangulate and enrich our understanding of wildfire-related health concerns.

\section{Conclusion}\label{sec6}

In this study, we employed an advanced topic modeling framework enhanced with large language models (LLMs) and human-in-the-loop (HITL) refinement to analyze social media discourse during the 2025 Los Angeles wildfires. Leveraging a dual strategy that combines Latent Dirichlet Allocation (LDA) and BERTopic models, along with multi-hierarchical and multi-label classification, we conducted comprehensive temporal trend and spatial pattern analyses. This approach systematically identified key situational awareness topics and crisis narratives, with particular emphasis on public health and mental health concerns.

Our study highlights the potential of underexplored social media as a powerful source for understanding, empathizing with, and assisting during immediate disaster responses. The boundless nature of social media facilitates real-time communication, resource coordination, and emotional expression, and therefore enables communities to organize, respond, and recover in a timely manner. Online communication reveals situational public health challenges, while virtual communities foster collective resilience through shared experiences and mutual support. By examining both situational awareness and crisis narratives, we can understand what is neglected during the response phase, which can help drive future mitigation initiatives that are more equitable, timely, and community centered. Furthermore, the dataset and analytical framework in our study are both scalable transferable, and they can contribute as a foundation for future research on the 2025 Los Angeles wildfires as well as other comparable disaster events.

\section*{Supplementary Information}

\renewcommand{\thefigure}{S\arabic{figure}}
\setcounter{figure}{0}

\begin{figure}[H]
\centering
\includegraphics[width=.9\textwidth]{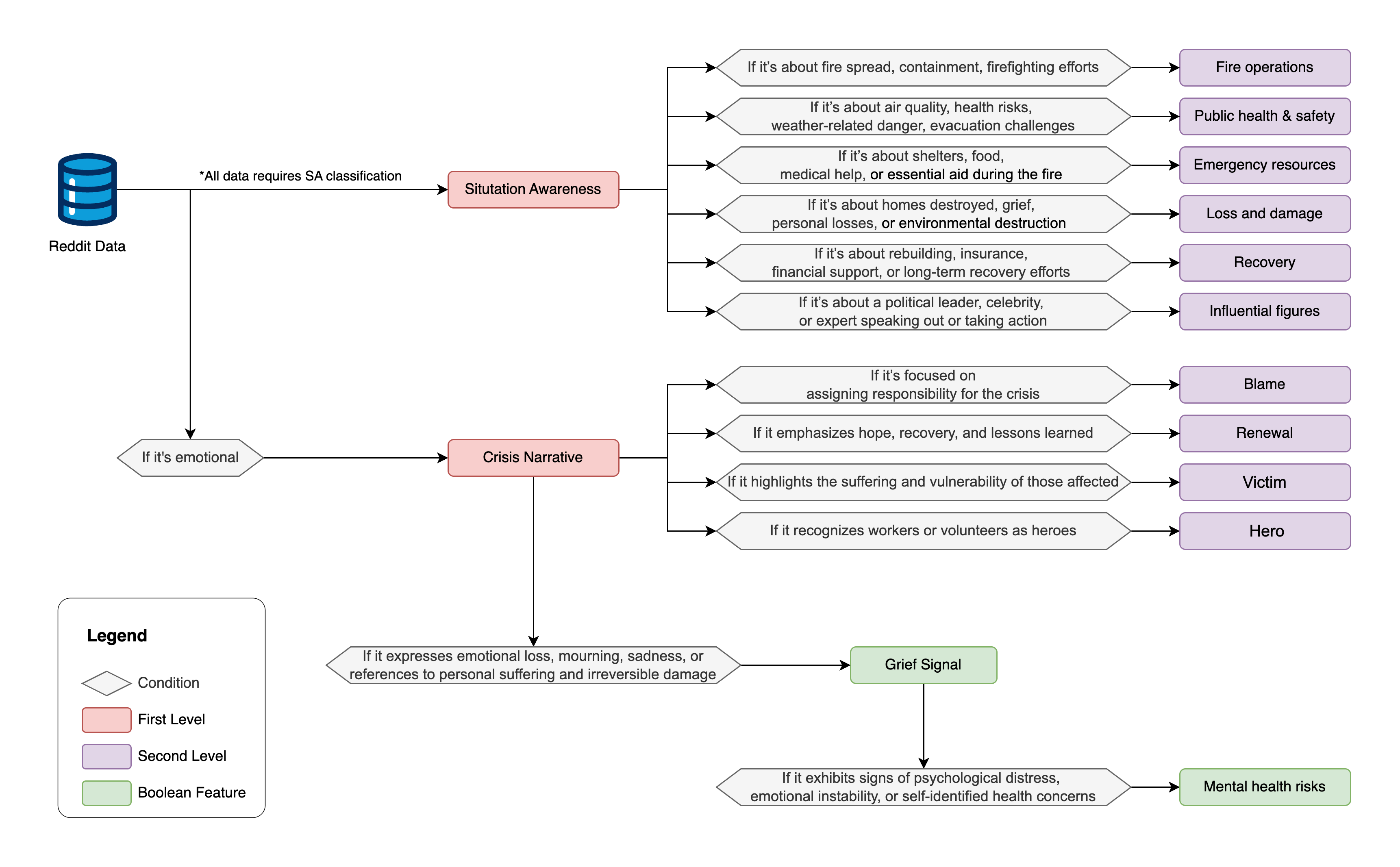}
\captionsetup{labelformat=empty}
\caption{\textbf{Supplementary Figure S1}. Decision tree based guideline for annotation flowchart.}\label{d_tree}
\end{figure}

\begin{figure}[H]
\centering
\includegraphics[width=.9\textwidth]{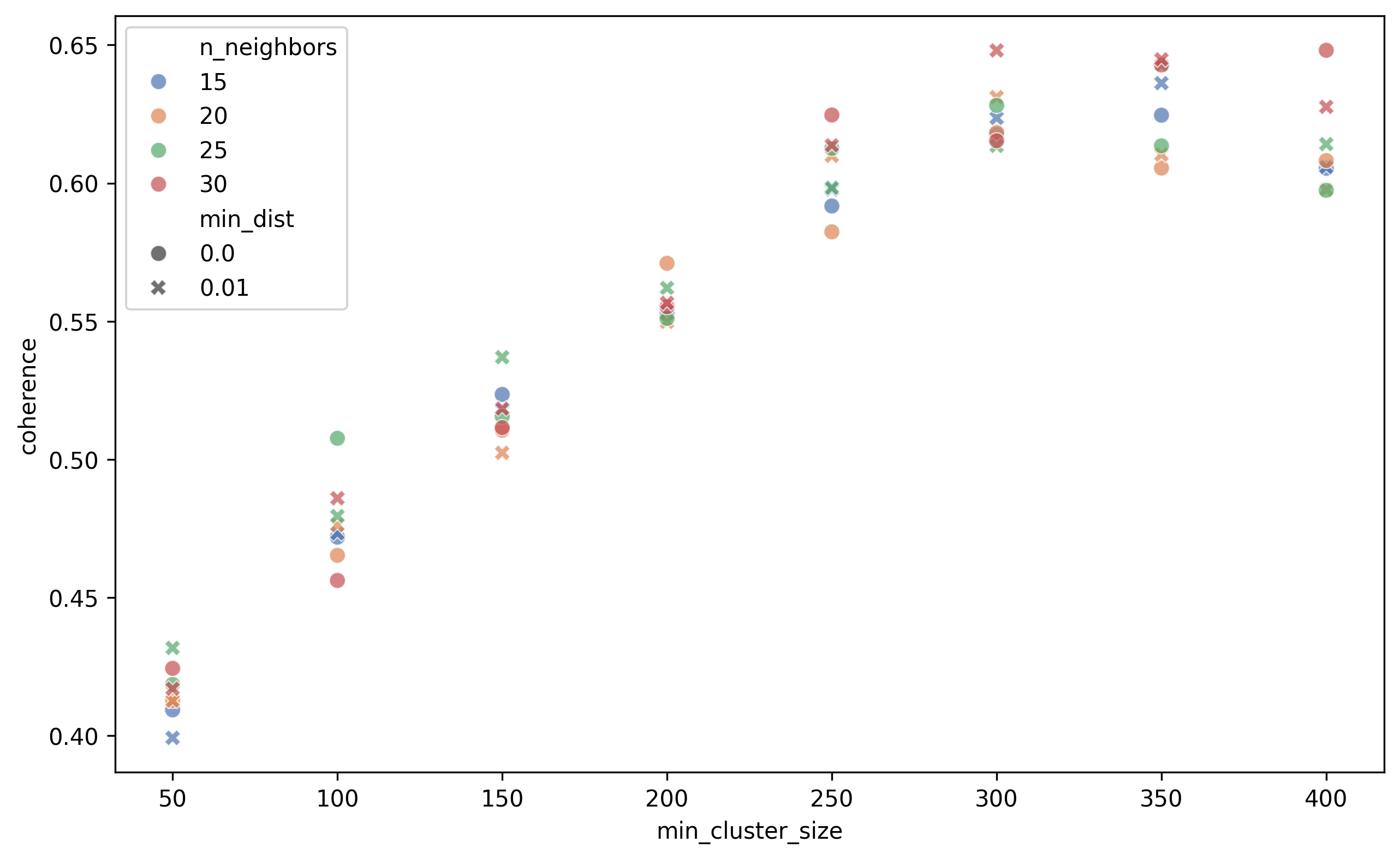}
\captionsetup{labelformat=empty}
\caption{\textbf{Supplementary Figure S2}. Parameter tuning results for UMAP and HDBSCAN.}\label{sf_parameter}
\end{figure}

\begin{figure}[H]
\begin{mdframed}[backgroundcolor=gray!10, linewidth=1pt, leftmargin=0pt, innerleftmargin=10pt, innerrightmargin=10pt]
\noindent
\begin{minipage}{\textwidth}
\textbf{Topic 3:} Air Quality and Health Concerns Post-Wildfires

\textbf{SA:} Public health and safety, Emergency resources, Recovery, Loss and damage

\textbf{CN:} Victim, Blame, Renewal

\textbf{Grief Signal:} Yes \quad \textbf{Mental Health Risk:} Yes

\vspace{0.3cm}

\textbf{Topic 24:} Evacuation Strategies: Cars vs. Buses in Emergencies

\textbf{SA:} Public health and safety, Emergency resources, Loss and damage

\textbf{CN:} Blame, Victim

\textbf{Grief Signal:} No \quad \textbf{Mental Health Risk:} No
\end{minipage}
\end{mdframed}

\captionsetup{labelformat=empty}
\caption{\textbf{Supplementary Figure S3}. Examples of mapping latent topics to pre-defined schema.}\label{latent_mapping}

\end{figure}

\section*{Data and code availability}

All code used in this study is publicly available on GitHub ( \url{https://github.com/MoonSulong/2025WildfireCA}). The de-identified datasets generated and analyzed during the study (text only; all images were removed to protect user privacy) are hosted on Hugging Face (\url{https://huggingface.co/datasets/Dragmoon/2025CalifoniaWildfire}).





\begin{appendices}



\end{appendices}

\bibliographystyle{unsrtnat}
\bibliography{fire}%

\end{document}